\RequirePackage{fix-cm}
\documentclass[11pt,a4,epsf]{article}

\usepackage{amssymb,amsmath,algorithm}

\usepackage{float}

\usepackage{graphicx}
\DeclareGraphicsExtensions{.eps}
\usepackage{epsfig}
\usepackage{epstopdf}
\usepackage{amsmath}
\usepackage{amssymb}
\usepackage{subfig}
\usepackage{rotating}

\usepackage{makeidx,epsfig}
\usepackage[margin=1.0in]{geometry}
\usepackage{amssymb,amsfonts}
\usepackage{amsmath}
\usepackage{amsbsy}
\usepackage{verbatim}
\usepackage[bottom]{footmisc}
\usepackage{algorithm}
\usepackage{algorithmicx}
\usepackage{algpseudocode}
\usepackage{colortbl,xcolor,pgfplots}
\usepackage{algorithm,algorithmicx,algpseudocode}

\usepackage[latin1]{inputenc}
\usepackage[english]{babel}

\usepackage{makeidx}
\usepackage{capt-of}
\usepackage[T1]{fontenc}
\usepackage{amsfonts}
\usepackage{amscd}
\usepackage{amssymb}
\usepackage{tabularx}
\usepackage{amssymb,bezier,graphicx,}
\usepackage{times,amssymb}
\usepackage{graphicx}
\usepackage{color}
\usepackage{cancel}
\usepackage{fancybox}
\usepackage{fancyhdr}
\usepackage{hyperref}
\usepackage{tikz}
\usepackage{enumerate}
\usepackage{epstopdf}
\usepackage{array}
\usepackage{slashbox}
\usepackage{verbatim}
\usepackage{float}
\usepackage[colorinlistoftodos]{todonotes}
\usepackage{framed}
\usepackage{anysize}
\usepackage{pdfpages}
\usepackage{stackrel}
\usepackage{booktabs}
\usepackage[mathscr]{eucal}
\usepackage{bm}
\usepackage{multirow}
\usepackage{caption}
\usepackage{mathrsfs} 
\usepackage{xcolor}

\def\R{\mathbb R}

\newtheorem{Thm}{Theorem}

\newtheorem{Rem}{Remark}

\begin{document}

\title{Location and portfolio selection problems: A unified framework}

\author{Justo Puerto$^{1,2}$, Mois\'{e}s Rodr\'{i}guez-Madrena $^{1,2}$,
Andrea Scozzari$^3$\\
\\
\normalsize $^1$ Institute of Mathematics University of Seville (IMUS), Seville, Spain\\
\normalsize $^2$ Department of Statistics and OR, Universidad de Sevilla, 41012 Seville, Spain \normalsize \{puerto, madrena\}@us.es\\
\normalsize $^3$ Faculty of Economics, Universit\`{a} degli Studi Niccol\`{o} Cusano Roma, Italy \normalsize  andrea.scozzari@unicusano.it}

\maketitle


\begin{abstract}
Given a set of assets and an investment capital, the classical portfolio selection problem consists in determining the amount of capital to be invested in each asset in order to build the most profitable portfolio. The portfolio optimization problem is naturally modeled as a mean-risk bi-criteria optimization problem where the mean rate of return of the portfolio must be maximized whereas a given risk measure must be minimized. Several mathematical programming models and techniques have been presented in the literature in order to efficiently solve the portfolio problem. A relatively recent promising line of research is to exploit clustering information of an assets network in order to develop new portfolio optimization paradigms. In this paper we endow the assets network with a metric based on correlation coefficients between assets' returns, and show how classical location problems on networks can be used for clustering assets. In particular, by adding a new criterion to the portfolio selection problem based on an objective function of a classical location problem, we are able to measure the effect of clustering on the selected assets with respect to the non-selected ones. Most papers dealing with clustering and portfolio selection models solve these problems in two distinct steps: cluster first and then selection. The innovative contribution of this paper is that we propose a Mixed-Integer Linear Programming formulation for dealing with this problem in a unified phase. The effectiveness of our approach is validated reporting some preliminary computational experiments on some real financial dataset.
\end{abstract}
\vskip 10 pt
\noindent Keywords: Portfolio selection; Clustering; $p$-median problem on networks; Conditional Value at Risk; Multicretiria optimization.

\section{Introduction and motivation}\label{intro}

Portfolio selection problem is among most the widely studied problems in financial literature since the seminal paper of Markowitz in 1952 \cite{Markowitz1952}. He formalized the notion of diversification in investing and used the variance of asset prices as a proxy of risk. In spite of its success, the model proposed by Markowitz has received many criticisms, first and foremost due to the large estimation errors on the vector of expected returns and on the covariance matrix \cite{ChopraZiemba_2013,Job_Kork1980,Merton_1980}. Hence, different directions have been proposed in the literature exploiting several strategies that consider improved estimates of correlation coefficients \cite{EltGruSpi2006}, alternative risk measures \cite{ArtDelEberHea_1999,BellDiber_2017}, effective diversification approaches (see, e.g., \cite{CeTar2017,DeMiGar2007,MaiRonTel_2010} and the references therein). As a consequence, new mathematical programming models and techniques were proposed in the literature in order to efficiently solve the portfolio selection problems incorporating the above mentioned different strategies (the interested reader can refer to \cite{ManOgrySpe_2015}).

At the early 2000s, some authors started in tackling the asset allocation problem via graph theory in which assets are identified by vertices of a graph $G$ and distances $d_{ij}$ assigned to pair of assets $i$ and $j$ incorporate the dependency structure of returns. Mantegna \cite{Mantegna_1999} was one of the first to construct asset graphs based on stock price correlations in order to detect the hierarchical organization inside a stock market. He showed that a minimal spanning tree of $G$ provides an arrangement of stocks which selects the most relevant connections among them. Moreover, the minimal spanning tree provides, in a direct way, the subdominant ultrametric hierarchical organization of the assets in a market. Based on this idea, Onnela et al. \cite{Onnela_et_al_2004} extended this approach for finding a subgraph of $G$, and, to this aim, they present a simple greedy-like procedure based on edge distances. Finally, they study the properties of the resulting subgraph such as number and size of clusters and clustering coefficient. These properties, based on real world dataset, were then compared against those of random graphs.

A relatively recent promising line of research is to exploit clustering information of an asset network in order to provide new portfolio optimization techniques. These approaches construct a portfolio by solving the classical Markowitz model by substituting the original correlation matrix with a correlation based clustering ultrametric matrix. The aim is twofold: (i) correlation based clustering may be seen as a filtering procedure and (ii) the portfolios selected by clustering algorithms are quite robust with respect to measurement noise due to the finiteness of sample size \cite{BerBru_2014,ClemGraHitaj2019,Tola_et_al_2008}. In particular, in \cite{Tola_et_al_2008} the authors substitute the original correlation matrix formed by $\frac{n(n-1)}{2}$ distinct elements with a smaller one formed by $n-1$ elements that represents a measure of distance (or similarity) between clusters of stocks. They apply the single linkage or the average linkage clustering procedure starting from initial distances $d_{ij}$ between assets $i$ and $j$. In \cite{ClemGraHitaj2019} the authors construct the starting $d_{ij}$ by using the Pearson Correlation measure, the Kendall rank correlation coefficient and the lower tail dependencies between each pair of assets.

\noindent To measure the level of interconnections of an asset with the whole system, in \cite{ClemGraHitaj2019} the authors also use the clustering coefficient introduced by Watts and Strogatz in \cite{WattsStrogatz_1998}, which refers to the number of existing triangles around a vertex with respect to the number of potential ones. Finally, they obtain a positive definite matrix $C$ of interconnections between pair of assets that is used in Markowitz's model in place of the original correlation matrix.

Other authors used graph-theoretic concepts to explore the global properties of stock markets by analyzing the structure of the underlying graph. In \cite{BogButeShiTruLaf_2014} the authors solve different classical NP-hard optimization problems to analyze the dependencies among stocks. A maximum clique problem was solved for detecting large clusters of similar and dissimilar
stocks according to the natural criterion of pairwise correlation. Independent sets, which represent groups of vertices with no connections, are sought for in the stock graph to find well-diversified portfolios.

The common feature of the above strategies is that they are based on two different and independent phases: clustering first and then analyze the number and properties of the assets clustered this way. In fact, all these papers do not investigate how the clustering phase could be used as an effective tool in the portfolio selection process for assessing portfolios' optimal weights.

The main contribution of this paper is to present a clustering and portfolio selection strategy in a unified framework that, in principle, could be used and adapted for finding portfolios minimizing different coherent risk measures usually considered in portfolio optimization. In this paper we model the clustering problem as a network $p$-median problem. This is a classical network facility location problem in the Operations Research literature. It was introduced in \cite{KarHak_1979} and, given an $n$ vertex graph $G$, the problem consists in finding a set of exactly $p$ vertices (facilities) in order to minimize the sum of the distances (accessibility criterion) between the selected facilities and the other vertices of the graph. It is shown in \cite{KarHak_1979} that the problem of finding a $p$-median of a network is NP-hard even when the network has a simple structure.

The $p$-median problem may be also interpreted in terms of cluster analysis, and the data clustering application is straightforward since it simply requires reinterpreting the medians as the median of a cluster instead of as a facility. In fact, in \cite{Hansen_et_al_2009} the authors observed that when squared Euclidean distance is used, the problem becomes a discretized version of the well-known $k$-means problem in cluster analysis, and the optimal solution to the $p$-median problem results in a partition of the vertex set into homogeneous clusters. Ng and Han \cite{NgHan_1994} showed that the $p$-median model is useful for detecting patterns and mining data as well as clustering. Thus, the $p$-median model can be a powerful tool for data mining applications. We also mention the research of Benati \cite{Benati_2008}, Benati and Garc\'{i}a \cite{BenGar_2014}, and Benati et al. \cite{BeGaPu_2018} on clustering based on $p$-median models.

In this paper we provide a novel mathematical programming formulation for the problem of simultaneously locating a set of $p$ facilities in a network of stocks while selecting a subset of assets minimizing a given risk measure. In particular, here we consider the Conditional Value at Risk to measure the riskiness of an investment. We note that an asset selected as a facility can be considered as a \emph{representative} of a cluster of stocks centered in it. We also show that our formulations are \emph{portable} in the sense that there is a subset of constraints that remains unchanged regardless of the risk measure considered. We test our formulations on different real world financial dataset, and compare the results of the different formulations from a profitability point of view.

The paper is organized as follows. In Section 2 we introduce some necessary definitions and notation along with an analysis of the NP-completeness of our unified problem in some special classes of graphs. In section 3 we provide the mathematical programming formulation of our problem, in addition to a description on how to compute some parameters of the model. Section 4 is devoted to the computational experiments that show the effectiveness of our approach on some real financial datasets. Finally, in the concluding section some remarks and future research are depicted.

\section{Notation, definitions and properties}

Consider a finite, connected and undirected graph with no self-loops $G=(V,E)$. We denote by $V(G)$ and $E(G)$ the vertex and edge set, respectively, with $|V(G)| = n$ and $|E(G)| = m$. An edge $e\in E(G)$ is identified by a pair of vertices $(i,j)$, with $i,j\in V(G)$. Suppose that a nonnegative real length $d(e)=d_{ij}$ is assigned to each edge $e\in E(G)$. Each edge is assumed to be rectifiable, so that $A(G)$ is the continuum set of points on the edges of $G$. We refer to interior points on an edge by their distances (along the edge) from the two end vertices of the edge. For any pair of points $x,y\in A(G)$, we let $d_{xy}$ denote the length of a shortest path in $A(G)$ connecting $x$ and $y$. We refer to $A(G)$ as the metric space induced by $G$ and the edge lengths. Consider a closed subset $X_p$ of $p$ points of $A(G)$; the distance $d_i(X_p)$ between a vertex $i$ and $X_p$ is defined as the smallest of the distances from $i$ to the points in $X_p$, that is: $d_i(X_p) =\min\{d_{ix}|x\in X_p\}$. Thus, we define:

$$F(X_p)=\sum\limits_{i\in V(G)}d_i(X_p),$$

\noindent the \emph{sum of the distances} of all the vertices in $G$ to the set $X_p$. The $p$-median problem on $G$ consists of finding a set $X^*_p$ such that

$$F(X^*_p)=\min\limits_{X_p\subset A(G)} \{F(X_p)\}.$$

The set $X^*_p$ is called a \emph{p-median} of $G$. It is well-known that there exists a $p$-median whose points are vertices, that is, $V^*_p=X^*_p$, with $V^*_p\subset V(G)$ \cite{KarHak_1979}, so that a \emph{p-median} of $G$ is a set $V^*_p$ such that

$$F(V^*_p)=\min\limits_{V_p\subset V(G)} \{F(V_p)\}.$$

\noindent The portfolio optimization method we propose here is a clustering approach based on the $p$-median optimization problem to select portfolios by modeling the stock market as a network. In other words, we construct a portfolio by selecting each asset from the $p$ clusters obtained by partitioning the (possibly complete) asset network $G=(V,E)$, where $V(G)$ is the set of stocks and the distances $d_{ij}$, $i,j\in V(G)$, are based on correlation coefficients $\rho_{ij}$ between each pair of assets.

\noindent Following the seminal paper by Markowitz \cite{Markowitz1952}, the classical portfolio selection problem consists in determining the amount of capital to be invested in each asset of a given stock market. The problem is modeled as a bi-criteria mean-risk optimization problem as follows:

$$\max\{[\mu(\mathbf{x}), -\varrho(\mathbf{x})]| \mathbf{x}\in \Delta\},$$

\noindent where $\Delta=\{\mathbf{x}\in \R^n| \mathbf{x}\geq 0, \sum\limits_{i=1}^n x_i = 1\}$, $x_i$ is the proportion of capital invested in asset $i$, $i=1,\ldots,n$, $\mu(\mathbf{x})$ is the expected rate of return of a portfolio $\mathbf{x}$, and $\varrho: \R^n \rightarrow \R$ is a (coherent) risk measure.

\noindent In this paper we show how the objective function of the classical $p$-median problem on networks can be used in order to filter the relevant information in a multivariate set of data. In particular, by means of an objective function of a classical location problem we are able to measure the clustering effect on the selected assets w.r.t. the non-selected ones. To this aim, given a network of assets $G=(V,E)$, we propose a tri-criteria optimization problem of the form:

$$\max\{[\mu(\mathbf{x}), -\varrho(\mathbf{x}), -F_p(\mathbf{x})]| \mathbf{x}\in \Delta, \; V_p\subset V(G)\},$$

\noindent being ${F_p(\mathbf{x})= F(V_p)}$. We note that we impose that a $p$-median of $G$ must be a subset $V_p$ of vertices (assets) of $G$ and a vertex $j\in V_p$ can be considered as the \emph{representative} asset of a given cluster. A vertex $i\notin V_p$ which is assigned to a vertex $j\in V_p$, is said to be represented by $j$.

We provide a multi-objective Mixed-Integer Linear Programming (MILP) formulation for dealing with the above problem, that can be written as follows:

\begin{subequations}
\begin{align}
\max & \,\,\,\,\, \displaystyle  \mu(\mathbf{x}) \noindent \label{B1}\\
\min & \,\,\,\,\, \displaystyle  \varrho(\mathbf{x}) \noindent \label{B2}\\
\min & \,\,\,\,\, \displaystyle  F_p(\mathbf{x}) = \sum\limits_{i=1}^n \sum_{j=1}^n d_{ij} z_{ij} \noindent \label{B3}\\
\mbox{ s.t. } &  \,\,\,\,\, \displaystyle  \mathbf{x}\in \Delta \noindent \label{B10}\\
&  \,\,\,\,\, \sum\limits_{j=1}^n z_{jj} = p  \noindent \label{B12}\\
&  \,\,\,\,\, \sum\limits_{j=1}^n z_{ij} = 1 \quad i=1,\ldots,n  \noindent \label{B13}\\
&  \,\,\,\,\, z_{ij} \leq z_{jj} \quad i,j=1,\ldots,n  \noindent \label{B14}\\
\nonumber \\
&  \,\,\,\,\, \ell_j z_{jj} \leq x_j\leq u_j z_{jj} \qquad j=1,\ldots,n  \noindent \label{B16}\\
\nonumber \\
&  \,\,\,\,\, z_{ij}\in \{0,1\} \quad i,j=1,\ldots,n \noindent \label{B15}
\end{align}
\end{subequations}

\noindent where:\\

$z_{jj} = \begin{cases} 1 \qquad \hbox{if asset $j$ is selected in the portfolio and considered as a representative}\\ 0 \qquad \hbox{otherwise}\end{cases}$

\vskip 8 pt

$z_{ij} = \begin{cases} 1 \qquad \hbox{if asset $i$ is not selected in the portfolio but it is represented by asset $j$}\\ 0 \qquad \hbox{otherwise.}\end{cases}$

\vskip 8 pt

\noindent Constraint (\ref{B12}) assures to select exactly $p$ representatives (facilities); constraints (\ref{B13}) guarantee that each asset belongs to exactly one cluster, whereas constraints (\ref{B14}) assure that an asset $i$ is represented by $j$ only if $j$ is selected as a representative of a cluster. Constraints (\ref{B16}) tie together the location and portfolio selection problems. In fact, constraint (\ref{B16}) states that if an asset $j$ is selected as a representative, then we invest in asset $x_j$ an amount of capital that must be at least equal to $\ell_j\geq 0$, to prevent holding of a position below a minimal allowable size, and at most $u_j\geq 0$, to prevent holding too large positions. In the following $\Gamma$ denotes the feasible region determined by (\ref{B10})-(\ref{B15}).
\vskip 8 pt

There exist different approaches for handling the multi-objective optimization problem (\ref{B1})-(\ref{B15}). Most of the methods  convert the given multi-objective program to a single-objective one after a scalarization of the different objective functions.
Among the most effective approaches one can consider the $\epsilon$-constraint method. It consists in solving the problem with respect to one of its objective functions whereas the remaining objective functions are included as constraints parametrically varying their right-hand-sides with values \cite{Ehrgott_2005}. Taking values $\varrho_0$, $\mu_0$, for the risk measure ($\varrho$) and the average return, respectively, our problem becomes:

\begin{subequations}
\begin{align}
\min & \,\,\,\,\, \displaystyle  \sum\limits_{i=1}^n \sum_{j=1}^n d_{ij} z_{ij} \noindent \label{S1}\\
\mbox{ s.t. } &  \,\,\,\,\, \displaystyle  (\mathbf{x},\mathbf{z})\in \Gamma \noindent \label{S2}\\
& \,\,\,\,\,  \varrho(\mathbf{x}) \le \varrho_0, \noindent \label{S3} \\
& \,\,\,\,\, \mu(\mathbf{x}) \ge \mu_0, \noindent \label{S4}.
\end{align}
\end{subequations}

\vskip 8 pt

\noindent This approach can also handle problems with integrality constraints by adequately choosing the different values of the right-hand-sides. From a portfolio selection viewpoint, as also shown in the seminal paper of Chang et al. \cite{ChangMeadeBeasley_2000}, the presence of integrality constraints (e.g., cardinality constraints) can make the efficient frontier discontinuous, where the discontinuities might imply that there are certain returns which no rational investor would consider.

\subsection{Complexity issues}

To analyze the complexity status of our location and portfolio selection problem, let us consider precisely the single-objective program (\ref{S1})-(\ref{S4}). It clearly contains the classical $p$-median problem on networks as a special case. Assume $\varrho_0= +\infty$ and $\mu_0=-\infty$, $\ell_j=0$ and $u_j=1$, then constraints (\ref{S3}) and (\ref{S4}) can be ignored, i.e., a feasible portfolio can be always provided by setting $x_j=\frac{1}{p}$, for all $j$ such that $z_{jj}=1$, and problem (\ref{S1})-(\ref{S4}), in fact, reduces to finding $p$ facilities in order to minimize the sum of the distances between the opened facilities and the vertices of $G$.

It is well known that the $p$-median problem is NP-hard even in the case where the network is a planar graph of maximum vertex degree 3, all of whose edges and vertices have weight 1 \cite{KarHak_1979}. The proof in \cite{KarHak_1979} is based on the fact that the $p$-median problem is polynomially equivalent to the NP-complete problem of deciding if there exists a dominating set of cardinality $p$ in a planar graph of maximum vertex degree 3 \cite{GareyJohnson_1979}. Later, in \cite{CornPerl_1984,CornStewart_1990} the authors show that the dominating set problem is NP-complete on some special classes of perfect graphs namely, comparability graphs, bipartite graphs, chordal graphs, split graphs, $k$-trees (arbitrary $k$), and undirected path graphs (which are defined as the intersection graphs of a set of undirected paths in a tree). Hence, the $p$-median problem remains NP-complete on the same classes of graphs. From the above discussion we have the following result.

\begin{Thm}
The location and portfolio selection problem (\ref{S1})-(\ref{S4}) is NP-hard for:
\begin{itemize}
\item[(i)] comparability graphs.
\item[(ii)] bipartite graphs.
\item[(iii)] chordal graphs.
\item[(iv)] split graphs.
\item[(v)] $k$-trees (arbitrary $k$).
\item[(vi)] undirected path graphs.
\end{itemize}
\end{Thm}

\noindent Finally, we observe that in \cite{CornStewart_1990} the authors also report the relationships among some families of perfect graphs. Therefore, the NP-hardness results imply a number of other NP-hardness results because of the containment relationships among the various perfect graph families.

\section{Solution method}

In this section we present the solution technique used for solving the multi-objective optimization problem (\ref{B1})-(\ref{B15}). We adopt the well-known $\epsilon$-constraint method which is based on retaining only one of the three objectives and to turn all the others into constraints \cite{Ehrgott_2005}. We retain the risk measure as objective, so that model (\ref{B1})-(\ref{B15}) can be rewritten as follows:

\begin{subequations}
\begin{align}
\min & \,\,\,\,\, \displaystyle  \varrho(\mathbf{x})\noindent \label{E1}\\
\mbox{ s.t. } &  \,\,\,\,\, \mu(\mathbf{x})\geq \mu_0 \noindent \label{E2}\\
& \,\,\,\,\, F_p(\mathbf{x})\leq F^0_p \noindent \label{E3}\\
&  \,\,\,\,\, \displaystyle  (\mathbf{x},\mathbf{z})\in \Gamma \noindent \label{E4}
\end{align}
\end{subequations}

\noindent where $\mu_0$ and $F_p^0$ are, respectively, a lower bound for $\mu(\mathbf{x})$ and an upper bound for $F_p(\mathbf{x})$.

\vskip 8 pt

\noindent Problem (\ref{E1})-(\ref{E4}) is a general \emph{location and portfolio selection} problem in the sense that any (coherent) risk measures can be considered as objective function. Nevertheless, in this paper we focus on some measures based only on the historical returns. In particular, we consider the Conditional Value at Risk (CVaR) as objective function \cite{RockafellarUryasev_2002}. We observe that other historical returns based measures could be used (Mean Absolute Deviation, Gini's index, minimax objective function...). There are several reasons for this choice. From a computational viewpoint, risk measures based on past observation periods lead to integer linear programming problems instead of nonlinear (e.g., quadratic) integer programs that are more difficult to solve. Moreover, in the case of the variance (or a correlation based) risk measure, it is well-known that it is based on the assumption that investors are risk averse and that either the distribution of the rate of return is multivariate normal or the utility of the investors is a quadratic function of the rate of returns. Unfortunately, neither of the two above conditions hold in practice (among others, see, e.g., \cite{ChopraZiemba_2013,KonnoYamazaki_1991,ManOgrySpe_2007} and the references therein). Finally, when the portfolio objective function is based on the variance, it has been found that the composition of the optimal portfolio can be very sensitive to estimation errors in the expected returns of the underlying assets \cite{ChopraZiemba_2013,Job_Kork1980,Merton_1980}.

Despite the above reasons, it is not possible to completely ignore the correlation between the time series of returns of each asset that must be considered in order to quantify a measure of similarity between pairs of stocks. A widespread choice consists in quantifying the similarity between two assets with Pearson's correlation. Thus, the asset graph $G$ represents the correlation structure between stocks where the weights/distances associated to each edge refer to the linear correlation between them \cite{ClemGraHitaj2019}. Therefore, following a common practice in the literature about filtering procedures based on correlation clustering \cite{Tola_et_al_2008}, we assign to each edge $(i,j)\in E(G)$ a similarity measure (distance) $d_{ij}=\sqrt{2(1-\rho_{ij})}$ where $\rho_{ij}=\frac{\sigma_{ij}}{\sqrt{\sigma_{ii}\sigma_{jj}}}$ is the correlation coefficient between assets' returns. We remark here that any alternative measures like the Kendall rank correlation coefficient or the Tail coefficient could be used as distances between assets (see, e.g., \cite{ClemGraHitaj2019}).
Hence, with these choices, in our model the two main criteria (location and risk) are evaluated with two different measures in order to make the clustering and the portfolio selection phases well characterized and distinct.

\vskip 8 pt

In the light of the above discussion, in the following we introduce problem (\ref{E1})-(\ref{E4}) with the CVaR as objective function. Let us consider $T$ different scenarios for the returns of the $n$ assets. Each scenario $t$ has associated a probability $p_t$, $t=1,\ldots,T$, with $\sum_{t=1}^T p_t =1$.
Let $r_{jt}$ be the rate of return of asset $j$ at time $t$, $j=1,\ldots,n$, $t=1,\ldots,T$. Let $\mu_j$ be the average rate of return of asset $j$, i.e., $\mu_j=\sum_{t=1}^T p_t r_{jt}$, $j=1,\ldots,n$. The rate of return at time $t$ of a portfolio $\mathbf{x}=(x_1,\ldots,x_n)\in \Delta$ is
$$ y_t(\mathbf{x}) = \sum_{j=1}^n r_{jt} x_j,$$
and the corresponding expected rate of return is
$$ \mu(\mathbf{x}) = \sum_{j=1}^n \mu_{j} x_j.$$

The CVaR of a portfolio $\mathbf{x}=(x_1,\ldots,x_n)\in \Delta$ with tolerance level $\beta\in (0,1]$ is defined as
$$ M_{\beta}(\mathbf{x}) = \min_{u_t} \{\dfrac{1}{\beta} \sum_{t=1}^T y_t(\mathbf{x}) u_t | \sum_{t=1}^T u_t = \beta, 0\leq u_t\leq p_t \quad t=1,\ldots,T\}.$$

In order to obtain a linear formulation for model (\ref{E1})-(\ref{E4}), we consider the dual of problem $M_{\beta}(\mathbf{x})$, that is
$$ M_{\beta}(\mathbf{x}) = \max_{\eta,d_t^-} \{\eta-\dfrac{1}{\beta} \sum_{t=1}^T p_t d_t^- : d^-_t\geq \eta - y_t(\mathbf{x}), d_t^- \geq 0 \quad t=1,\ldots,T, \eta\in \mathbb{R}\}.$$
Hence our problem (\ref{E1})-(\ref{E4}) can be reformulated as a Mixed-Integer Linear Programming (MILP) problem
\begin{subequations}
\begin{align}
\max & \,\,\,\,\, \displaystyle   \eta-\dfrac{1}{\beta} \sum_{t=1}^T p_t d_t^- \noindent \label{D1}\\
\mbox{ s.t. } &  \,\,\,\,\, \displaystyle   d^-_t\geq \eta - y_t(\mathbf{x})\quad t=1,\ldots, T \noindent \label{D2}\\
&  \,\,\,\,\, \displaystyle   d_t^- \geq 0\quad t=1,\ldots, T \noindent \label{D3}\\
&  \,\,\,\,\, \displaystyle   \eta\in \mathbb{R} \noindent \label{D4}\\
&  \,\,\,\,\, \displaystyle   \mu(\mathbf{x})\geq \mu_0 \noindent \label{D5}\\
&  \,\,\,\,\, \displaystyle  F_p(\mathbf{x})\leq F_p^0 \noindent \label{D6}\\
&  \,\,\,\,\, \displaystyle  (\mathbf{x},\mathbf{z})\in \Gamma. \noindent \label{D7}
\end{align}
\end{subequations}

\begin{Rem}\label{Rem1}
\cite{GarLabMar_2011} In a $p$-median problem on networks, variable $z_{ij}$, $i\neq j$, takes value 1 in an optimal solution for a representative $j$ such that $d_{ij} = \min\limits_{k:z_{kk}=1} \{d_{ik}\}$.
\end{Rem}

\noindent Remark \ref{Rem1} implies that in model (\ref{D1})-(\ref{D7}) we can consistently reduce the number of binary variables from $O(n^2)$ to $O(n)$ by substituting (\ref{B15}) in $\Gamma$ with the following constraints

\begin{align}
z_{jj}\in \{0,1\} \quad j=1,\ldots,n \noindent  \nonumber \\
\nonumber \\
z_{ij}\geq 0 \quad 1\leq i,j \leq n,\; i\neq j.\noindent  \nonumber
\end{align}



\subsection{Selection of $F_p^0$}\label{subsection_3.1}

The selection of the lower bound on the portfolio expected return $\mu_0$ in model (\ref{D1})-(\ref{D7}) is, evidently, of the taste of the investor. However, how to control the effect of clusterization by means of the upper bound $F_p^0$ is not obvious. In this section we describe how to compute $F_p^0$ in (\ref{D1})-(\ref{D7}).
\vskip 8 pt

\noindent Consider the problem

\begin{subequations}
\begin{align}
\min & \,\,\,\,\, \displaystyle F_p(\mathbf{x})  \label{PMP_1}\\
\mbox{ s.t. } &  \,\,\,\,\, \mu(\mathbf{x})\geq \mu_0  \label{PMP_2}\\
&  \,\,\,\,\, \displaystyle  (\mathbf{x},\mathbf{z})\in \Gamma \label{PMP_3}
\end{align}
\end{subequations}
\vskip 8 pt

\noindent and let $F_p^{\ell}$ be its optimal value. Note that $F_p^{\ell}$ is the minimum value for $F_p^0$ that makes problem (\ref{D1})-(\ref{D7}) feasible.
\vskip 8 pt

\noindent Let $\mathcal{S}$ be the set of feasible solutions defined by (\ref{B10}) together with (\ref{D2})-(\ref{D4}). Consider now the problem
\begin{subequations}
\begin{align}
\max & \,\,\,\,\, \displaystyle  \eta-\dfrac{1}{\beta} \sum_{t=1}^T p_t d_t^- \label{LCVaR_1}\\
\mbox{ s.t. } & \,\,\,\,\, (\mathbf{x},\mathbf{d}^-,\eta)\in \mathcal{S} \label{LCVaR_2}\\
&  \,\,\,\,\, \displaystyle   \mu(\mathbf{x})\geq \mu_0 \noindent \label{LCVaR_3}\\
&  \,\,\,\,\, \sum\limits_{j=1}^n z_{j} = p  \noindent \label{LCVaR_4}\\
&  \,\,\,\,\, \ell_j z_{j} \leq x_j\leq u_j z_{j} \qquad j=1,\ldots,n  \noindent \label{LCVaR_5}\\
&  \,\,\,\,\, z_{j}\in \{0,1\} \qquad j=1,\ldots,n \noindent \label{LCVaR_6}
\end{align}
\end{subequations}
\vskip 8 pt

\noindent and let $X_p$ be the set of the $p$ selected assets in an optimal solution. Then, $F_p(X_p)=F_p^u$ is the tightest upper bound for $F_p^0$, that is, for all $F_p^0\geq F_p^u$ the objective $F_p(\mathbf{x})$ is negligible in (\ref{D1})-(\ref{D7}).

\vskip 8 pt
\noindent Finally, given $\gamma \in [0,1]$, the parameter $F_p^0$ can be computed as follows
$$F_p^0 = \gamma F_p^{\ell} + (1-\gamma)F_p^u. $$
In this way, the selection of the parameter $\gamma$ allows us to control the required clustering effect, from the highest one ($\gamma=1$) to the lowest one ($\gamma=0$).
\vskip 8 pt

\section{Experimental results}

This section presents an empirical analysis with the aim of evaluating the performance of the portfolios selected by our location and selection model (\ref{D1})-(\ref{D7}). In addition, we also compare our model with the pure CVaR approach \cite{RockafellarUryasev_2002}, regarded as the benchmark program, and the CVaR model with cardinality constraints.
\vskip 8 pt

\noindent The pure CVaR model is \cite{ManOgrySpe_2007,RockafellarUryasev_2002}

\begin{subequations}
\begin{align}
\max & \,\,\,\,\, \displaystyle  \eta-\dfrac{1}{\beta} \sum_{t=1}^T p_t d_t^- \label{CVaR_1}\\
\mbox{ s.t. } & \,\,\,\,\, (\mathbf{x},\mathbf{d}^-,\eta)\in \mathcal{S} \label{CVaR_2}\\
&  \,\,\,\,\, \displaystyle   \mu(\mathbf{x})\geq \mu_0 \noindent \label{CVaR_3}
\end{align}
\end{subequations}

\vskip 8 pt
\noindent and the CVaR model with a limited number of assets (CVaR-CC) is precisely the above model (\ref{LCVaR_1})-(\ref{LCVaR_6}) (see, e.g., \cite{CeScoTar_2015}).
\vskip 8 pt

\subsection{Data sets}
We test all the above portfolio selection strategies on some real-world dataset belonging to the major stock markets across the world. We consider the following datasets:

\begin{itemize}
\item[1.] DJIA (Dow Jones Industrial Average, USA), containing 28 assets and 1353 price observations (period: 07/05/1990 - 04/04/2016);
\item[2.] EUROSTOXX50 (Europe's leading blue-chip index, EU), containing 49 assets and 729 price observations (period: 22/04/2002-04/04/2016);
\item[3.] FTSE100 (Financial Times Stock Exchange, UK), containing 83 assets and 625 price observations (period: 19/04/2004-04/04/2016);
\item[4.] SP500 (Standard $\&$ Poor's, USA), containing 442 assets and 573 observations (period: 18/04/2005-04/04/2016).
\end{itemize}

Each dataset consists of weekly prices data, and, as observed in \cite{ManOgrySpe_2007}, ``the choice of weekly periodicity is consistent with the objective of reducing estimation errors \cite{Simaan_1997}''. To evaluate the performance of our models in practice, we divide the observations in two sets, where the first one is regarded as the past (in-sample window), and so it is known, and the rest is regarded as the future (out-of-sample window), supposed unknown at the time of portfolio selection. The in-sample window is used for selecting the portfolio, while the out-of-sample one is used for testing the performance of the selected portfolio. Let $0,1,\ldots,T$ be the observations in the in-sample window; for each dataset, to compute the $T\times n$ matrix of the historical returns, we consider the time series of the prices of the $n$ stocks and denote by $P_i(t)$ the price of the $i$th asset at time $t$, $t=0,\ldots,T$. The $i$th asset return at time $t$ is computed as $r_{it}=\frac{P_i(t)-P_i(t-1)}{P_i(t-1)}$, with $t=1,\ldots,T$. As is standard practice in the literature, to model the correlation structure between stocks we refer to the series of the in-sample logarithmic returns of each asset $i$ computed as $R_{it}=\ln P_i(t)- \ln P_i(t-1)$, with $t=1,\ldots,T$ (see, e.g., \cite{Mantegna_1999,Onnela_et_al_2004}). Then, we compute the distances $d_{ij}$ between assets $i$ and $j$, $i\ne j=1,\ldots,n$, as $d_{ij}=\sqrt{2(1-\rho_{ij})}$ where $\rho_{ij}$ is the Pearson correlation coefficient between assets' logarithmic returns. We do not set any threshold for the correlation coefficients, thus the resulting asset graph $G$ is a complete weighted graph.
\vskip 8 pt
\noindent In our experiments we use a \emph{rolling time window} scheme allowing for the possibility of rebalancing the portfolio composition during the holding period, at fixed intervals. Following \cite{JegTit_2001,ManOgrySpe_2007}, for each dataset we adopt a period of 104 weeks (two years) as in-sample window and, in a first experimental set-up, we consider 52 weeks (one year) as out-of-sample, with rebalancing allowed every 52 weeks. In this phase, for each value of $p$, each model solves overall 55 problems (24 for DJIA, 12 for EUROSTOXX50, 10 for FTSE100 and 9 for SP500).

\vskip 8 pt

\noindent As in several other papers (see, e.g., \cite{CeScoTar_2013} and the references therein) the values of $p$ for the number of representatives, that are also used for bounding the number of assets in a portfolio, are set to $p=5,10,15,20$. We choose $\ell_i=\frac{1}{n}$ and $u_i=1$ as lower and upper bounds for the minimum and maximum allowable amount of capital invested in each asset $i$, $i=1,\ldots,n$.

\vskip 8 pt

\noindent In portfolio selection problems, the out-of-sample performance of a portfolio is generally evaluated by using some performance measures. In our experiments we consider the following measures (see, e.g., \cite{BruCeScoTar_2017})

\begin{itemize}
\item [1.] \emph{Sharpe Ratio} (Sh) (\cite{Sharpe1966,Sharpe1994}): it is defined as the ratio between the average of the out-of-sample return of a portfolio $\mathbf{x}$, $\mu^{\rm{out}}(\mathbf{x})$, minus a constant risk free rate of return $r_f$ (that we set equal to 0), and its standard deviation, namely:
$$\frac{E[\mu^{\rm{out}}(\mathbf{x}) - r_f]}{\sigma(\mu^{\rm{out}}(\mathbf{x}))}.$$

\noindent The larger is the value of the index, the better is the portfolio performance.

\item[2.] \emph{Average return} (Av): it is defined as the average $E[\mu^{\rm{out}}(\mathbf{x})]$ of the out-of-sample returns of a portfolio.

\end{itemize}

The models have been implemented in MATLAB R2017A and they make calls to XPRESS solver version 8.5 for solving the MILP programs. All experiments were run in a computer DellT5500 with a processor Intel(R) Xeon(R) with a CPU X5690 at 3.75 GHz and 48 GB of RAM memory.
\vskip 8 pt
\noindent We report the out-of-sample performances in the following tables. In each table, the first column refers to the models implemented for the comparisons. In particular, with $\gamma=0.1,\ldots,1$ we refer to our MILP model (\ref{D1})-(\ref{D7}) for the different values of $\gamma$ used for finding the parameter $F_p^0$ which bounds the function $F_p(\mathbf{x})$ in (\ref{D1})-(\ref{D7}) (see Section \ref{subsection_3.1}). CVaR-CC and CVaR refer to the cardinality constraint CVaR model (\ref{LCVaR_1})-(\ref{LCVaR_6}) and to the pure CVaR model (\ref{CVaR_1})-(\ref{CVaR_3}), respectively. In all the models, we choose the average index return in each in-sample period as the lower bound $\mu_0$ on the expected rate of return. We also consider the out-of-sample performance of the market index (Index) as benchmark. For each value of $p=5,10,15,20$, in the columns we report the out-of-sample average return (Av-$p$) and the value of the Sharpe Ratio (Sh-$p$). In addition, for the first three dataset (DJIA, EUROSTOXX50 and FTSE100) we also report the average in-sample solution times. Finally, for the pure CVaR model, we also report the average number of assets in the optimal in-sample portfolios.

\noindent For the SP500 dataset, due to the size of the corresponding MILP model, we were unable to find an optimal solution in reasonable times. Thus, we set a time limit of 7200 seconds, and in the corresponding table we report the average percentage GAP (GAP$\%$) in place of the solution times. Furthermore, we adopt a special methodology for solving the model in each in-sample period. Note that MILP model (\ref{D1})-(\ref{D7}) for $\gamma=1$ can be solved in two easy steps: first, solve problem (\ref{PMP_1})-(\ref{PMP_3}) and then, after having fixed the binary variables $z_j$ according to the $p$-median solution found, solve problem (\ref{LCVaR_1})-(\ref{LCVaR_3}). Note also that the optimal solution of problem (\ref{D1})-(\ref{D7}) for $\gamma=1$ is a feasible solution of problem (\ref{D1})-(\ref{D7}) for $\gamma=0.9$, thus it can be used as initial solution in the branching procedure in order to get better solutions in smaller times. The same applies for a feasible solution of problem (\ref{D1})-(\ref{D7}) for $\gamma=0.9$ and problem (\ref{D1})-(\ref{D7}) for $\gamma=0.8$, and so on.

\vskip 8 pt
\noindent In the tables, in bold we provide the best values of the out-of-sample performance measures for each value of $p$. In order to highlight the effectiveness of our MILP approach, for each $p$ and value of $\gamma$, we write the values of the two performance indexes in \emph{italic} if they are both better than the corresponding values of the other models (CVaR-CC, CVaR, Index). In this way we can have a count of how many times our approach outperforms the other competing portfolio selection models.

\begin{sidewaystable}
\caption{Out-of-sample performances for DJIA with a 52 weeks rebalancing}
\label{Table_1}
\centering
\small
\begin{tabular}{|c| |c|c|c|  |c|c|c|  |c|c|c|  |c|c|c|}
\toprule
   Model& Av-5 ($\cdot 10^{-3}$) & Sh-5  & Time-5 & Av-10 ($\cdot 10^{-3}$) & Sh-10 & Time-10 & Av-15 ($\cdot 10^{-3}$) & Sh-15 & Time-15 & Av-20 ($\cdot 10^{-3}$) & Sh-20 & Time-20 \\
  \hline
  \midrule
$\gamma$=0.1	&	2.064	&	0.091	&	0.587	&	1.848	&	0.087	&	0.620	&	\emph{1.903}	&	\emph{0.089}	&	 0.566	&	\emph{2.162}	 &	\emph{0.101}	&	0.907\\
$\gamma$=0.2	&	1.927	&	0.083	&	0.536	&	\emph{1.882}	&	\emph{0.089}	&	0.628	&	\emph{1.997}	&	 \emph{0.094}	&	 1.144	&	\emph{2.178}	&	\emph{0.102}	 &	0.894\\
$\gamma$=0.3	&	2.088	&	0.091	&	0.718	&	1.820	&	0.086	&	0.597	&	\emph{1.892}	&	\emph{0.090}	&	 0.834	&	\emph{2.163}	 &	\emph{0.101}	&	1.540\\
$\gamma$=0.4	&	2.206	&	0.097	&	0.693	&	1.770	&	0.084	&	0.986	&	\emph{2.034}	&	\emph{0.096}	&	 1.061	&	\emph{2.225}	 &	\emph{0.104}	&	1.173\\
$\gamma$=0.5	&	2.333	&	0.103	&	0.642	&	1.820	&	0.084	&	0.779	&	\emph{2.001}	&	\emph{0.094}	&	 0.722	&	\emph{2.235}	 &	\emph{0.104}	&	1.306\\
$\gamma$=0.6	& \emph{2.338}	&	\emph{0.104}	&	0.630	&	\emph{1.913}	&	\emph{0.088}	&	0.741	&	\emph{1.963}	 &	 \emph{0.092}	&	1.083	&	\emph{2.166}	&	 \emph{0.101}	&	1.179\\
$\gamma$=0.7	&	\textbf{\emph{2.366}}	&	\textbf{\emph{0.105}}	&	0.685	&	\emph{1.985}	&	\emph{0.091}	&	0.731	 &	\textbf{\emph{2.100}}	&	\textbf{\emph{0.099}}	&	 0.949	&	\emph{2.122}	&	\emph{0.099}	&	0.981\\
$\gamma$=0.8	&	1.933	&	0.086	&	0.654	&	\emph{1.989}	&	\emph{0.091}	&	0.623	&	\emph{2.068}	&	 \emph{0.097}	&	1.044	&	\emph{2.276}	&	\emph{0.106}	 &	1.177\\
$\gamma$=0.9	&	2.285	&	0.099	&	0.667	&	\emph{2.006}	&	\emph{0.091}	&	0.680	&	\emph{2.060}	&	 \emph{0.097}	&	0.945	&	\emph{2.157}	&	\emph{0.100}	 &	1.670\\
$\gamma$=1	&	2.179	&	0.088	&	0.082	&	\textbf{\emph{2.321}}	&	\textbf{\emph{0.105}}	&	0.130	&	 \emph{2.098}	 &	\emph{0.097}	&	0.191	&	\textbf{\emph{2.379}}	 &	\textbf{\emph{0.108}}	&	0.139\\
\hline
\hline
CVaR-CC		&	2.279	&	0.104	&	0.290	&	1.796	&	0.085	&	0.117	&	1.874	&	0.089	&	0.465	&	2.112	&	 0.098	&	0.425
\\
\hline
  \midrule
  & Av ($\cdot 10^{-3}$) & Sh & Time & n. of assets\\
    \cmidrule{1-5}
   CVaR		&	1.856	&	0.087	&	0.003	&	8.6		\\
   Index	&	1.608	&	0.068	&	-	    &	28		\\
     \cmidrule{1-5}
\end{tabular}
\end{sidewaystable}

\begin{sidewaystable}
\caption{Out-of-sample performances for EUROSTOXX50 with a 52 weeks rebalancing}
\label{Table_2}
\centering
\small
\begin{tabular}{|c| |c|c|c|  |c|c|c|  |c|c|c|  |c|c|c|}
\toprule
   Model& Av-5 ($\cdot 10^{-3}$) & Sh-5  & Time-5 & Av-10 ($\cdot 10^{-3}$) & Sh-10 & Time-10 & Av-15 ($\cdot 10^{-3}$) & Sh-15 & Time-15 & Av-20 ($\cdot 10^{-3}$) & Sh-20 & Time-20 \\
  \hline
  \midrule
$\gamma$=0.1	&	2.008	&	0.075	&	1.239	&	1.946	&	0.074	&	1.528	&	1.908	&	0.074	&	1.383	&	 \emph{2.020}	 &	\emph{0.077}	&	2.723	\\
$\gamma$=0.2	&	2.051	&	0.076	&	1.108	&	\emph{1.977}	&	\emph{0.075}	&	2.134	&	1.936	&	0.075	&	 1.888	&	1.973	 &	0.076	&	2.203	\\
$\gamma$=0.3	&	1.979	&	0.073	&	1.361	&	\emph{1.974}	&	\emph{0.074}	&	2.336	&	\emph{1.971}	&	 \emph{0.076}	&	3.167	&	1.993	 &	0.076	&	2.720	\\
$\gamma$=0.4	&	1.972	&	0.073	&	1.670	&	1.950	&	0.074	&	1.596	&	1.925	&	0.074	&	1.579	&	1.981	 &	0.075	&	2.915	\\
$\gamma$=0.5	&	1.771	&	0.063	&	1.845	&	\emph{2.047}	&	\emph{0.077}	&	1.610	&	\emph{2.037}	&	 \textbf{\emph{0.078}}	&	2.699	&	\emph{2.039}	 &	 \emph{0.078}	&	3.446	\\
$\gamma$=0.6	&	1.874	&	0.069	&	1.977	&	1.910	&	0.072	&	1.582	&	1.944	&	0.074	&	2.512	&	 \emph{2.054}	&	\emph{0.079}	&	3.067	\\
$\gamma$=0.7	&	1.990	&	0.073	&	2.137	&	\emph{2.005}	&	\emph{0.076}	&	2.145	&	\emph{2.006}	&	 \emph{0.076}	&	3.593	&	\emph{2.051}	 &	\emph{0.078}	 &	4.445	\\
$\gamma$=0.8	&	1.766	&	0.064	&	2.151	&	\textbf{\emph{2.090}}	&	\textbf{\emph{0.078}}	&	2.810	&	 \textbf{\emph{2.059}}	&	\textbf{\emph{0.078}}	&	2.770	&	 \emph{2.063}	 &	\emph{0.078}	&	3.749	\\
$\gamma$=0.9	&	1.714	&	0.062	&	2.068	&	\emph{1.979}	&	\emph{0.074}	&	2.134	&	\emph{2.045}	&	 \emph{0.076}	&	3.350	&	\textbf{\emph{2.111}}	 &	 \textbf{\emph{0.080}}	&	5.329	\\
$\gamma$=1	&	0.868	&	0.030	&	0.274	&	1.967	&	0.065	&	0.241	&	1.803	&	0.067	&	0.153	&	1.887	&	 0.070	&	0.285	\\
\hline
\hline
CVaR-CC		&	\textbf{\emph{2.103}}	&	\textbf{\emph{0.078}}	&	0.327	&	1.955	&	0.073	&	0.077	&	1.939	&	 0.075	&	0.102	&	2.008	&	 0.077	&	0.150	\\
\hline
  \midrule
  & Av ($\cdot 10^{-3}$) & Sh & Time & n. of assets\\
    \cmidrule{1-5}
   CVaR		&	1.922	&	0.072	&	0.002	&	7.2		\\
   Index	&	0.509	&	0.017	&	-	    &	49		\\
     \cmidrule{1-5}
\end{tabular}
\end{sidewaystable}

\begin{sidewaystable}
\caption{Out-of-sample performances for FTSE100 with a 52 weeks rebalancing}
\label{Table_3}
\centering
\small
\begin{tabular}{|c| |c|c|c|  |c|c|c|  |c|c|c|  |c|c|c|}
\toprule
   Model& Av-5 ($\cdot 10^{-3}$) & Sh-5  & Time-5 & Av-10 ($\cdot 10^{-3}$) & Sh-10 & Time-10 & Av-15 ($\cdot 10^{-3}$) & Sh-15 & Time-15 & Av-20 ($\cdot 10^{-3}$) & Sh-20 & Time-20 \\
  \hline
  \midrule
$\gamma$=0.1	&	2.066	&	0.089	&	17.495	&	\emph{1.581}	&	\emph{0.074}	&	6.434	&	1.542	&	0.072	&	 2.732	&	\emph{1.698}	 &	\emph{0.080}	&	2.309	\\
$\gamma$=0.2	&	1.830	&	0.079	&	20.030	&	\emph{1.687}	&	\emph{0.078}	&	8.577	&	1.593	&	0.075	&	 2.975	&	1.686	 &	0.079	&	2.546	\\
$\gamma$=0.3	&	1.627	&	0.071	&	24.037	&	\emph{1.596}	&	\emph{0.075}	&	24.238	&	1.540	&	0.072	&	 2.658	&	1.653	 &	0.078	&	3.724	\\
$\gamma$=0.4	&	1.807	&	0.079	&	29.242	&	\emph{1.623}	&	\emph{0.075}	&	14.249	&	\emph{1.639}	&	 \emph{0.077}	&	2.809	&	1.597	 &	0.076	&	2.605	\\
$\gamma$=0.5	&	\textbf{\emph{2.104}}	&	\textbf{\emph{0.092}}	&	27.456	&	\emph{1.683}	&	\emph{0.079}	&	 17.795	&	\emph{1.760}	&	\emph{0.082}	&	4.379	&	 1.635	 &	0.077	&	3.760	\\
$\gamma$=0.6	&	1.921	&	0.083	&	26.529	&	\textbf{\emph{1.870}}	&	\textbf{\emph{0.085}}	&	16.946	&	 \emph{1.698}	&	\emph{0.079}	&	6.225	&	\emph{1.699}	 &	 \emph{0.080}	&	3.393	\\
$\gamma$=0.7	&	1.628	&	0.066	&	22.059	&	\emph{1.681}	&	\emph{0.077}	&	9.477	&	1.590	&	0.074	&	 6.440	&	\emph{1.710}	 &	\emph{0.081}	&	2.966	\\
$\gamma$=0.8	&	1.310	&	0.053	&	19.084	&	\emph{1.808}	&	\emph{0.081}	&	13.817	&	\textbf{\emph{1.870}}	&	 \textbf{\emph{0.087}}	&	5.646	&	\emph{1.817}	 &	 \emph{0.085}	&	3.238	\\
$\gamma$=0.9	&	1.499	&	0.059	&	10.887	&	\emph{1.716}	&	\emph{0.077}	&	8.265	&	\emph{1.785}	&	 \emph{0.082}	&	5.811	&	\emph{1.845}	 &	\emph{0.086}	 &	3.873	\\
$\gamma$=1	&	0.925	&	0.026	&	2.271	&	1.480	&	0.057	&	0.873	&	1.754	&	0.069	&	1.261	&	 \textbf{\emph{2.113}}	&	 \textbf{\emph{0.095}}	&	0.791	\\
\hline
\hline
CVaR-CC		&	2.045	&	0.091	&	0.442	&	1.492	&	0.069	&	0.317	&	1.623	&	0.076	&	0.329	&	1.696	&	 0.080	&	0.225	
\\
\hline
  \midrule
  & Av ($\cdot 10^{-3}$) & Sh & Time & n. of assets\\
    \cmidrule{1-5}
   CVaR		&	1.528	&	0.071	&	0.008	&	11.7		\\
   Index	&	0.040	&	0.015	&	-	    &	83		\\
     \cmidrule{1-5}
\end{tabular}
\end{sidewaystable}

\begin{sidewaystable}
\caption{Out-of-sample performances for SP500 with a 52 weeks rebalancing}
\label{Table_4}
\centering
\small
\begin{tabular}{|c| |c|c|c|  |c|c|c|  |c|c|c|  |c|c|c|}
\toprule
   Model& Av-5 ($\cdot 10^{-3}$) & Sh-5  & GAP$\%$-5 & Av-10 ($\cdot 10^{-3}$) & Sh-10 & GAP$\%$-10 & Av-15 ($\cdot 10^{-3}$) & Sh-15 & GAP$\%$-15 & Av-20 ($\cdot 10^{-3}$) & Sh-20 & GAP$\%$-20 \\
  \hline
  \midrule
$\gamma$=0.1	&	1.970	&	0.080	&	23.634	&	2.452	&	0.103	&	4.371	&	\textbf{\emph{2.266}}	&	 \textbf{\emph{0.098}}	&	0.302	&	2.165	 &	0.095	&	0.058	\\
$\gamma$=0.2	&	1.877	&	0.075	&	24.092	&	2.466	&	0.104	&	4.814	&	\emph{2.255}	&	\textbf{\emph{0.098}}	 &	0.304	&	\emph{2.170}	 &	\emph{0.095}	&	 0.063	\\
$\gamma$=0.3	&	2.028	&	0.082	&	25.405	&	2.478	&	0.104	&	5.243	&	2.186	&	0.094	&	0.404	&	 \emph{2.170}	 &	\emph{0.095}	&	0.065	\\
$\gamma$=0.4	&	1.881	&	0.075	&	31.787	&	2.402	&	0.103	&	6.853	&	2.174	&	0.094	&	0.540	&	 \emph{2.181}	 &	\emph{0.095}	&	0.058	\\
$\gamma$=0.5	&	1.861	&	0.073	&	34.658	&	2.271	&	0.097	&	8.286	&	2.182	&	0.095	&	0.629	&	 \emph{2.193}	 &	\emph{0.096}	&	0.037	\\
$\gamma$=0.6	&	1.980	&	0.077	&	35.634	&	2.024	&	0.083	&	9.302	&	2.077	&	0.089	&	1.293	&	 \emph{2.183}	 &	\emph{0.095}	&	0.097	\\
$\gamma$=0.7	&	\textbf{\emph{2.386}}	&	\textbf{\emph{0.098}}	&	43.565	&	2.182	&	0.089	&	14.402	&	2.109	&	 0.090	&	2.527	&	\emph{2.216}	 &	\emph{0.096}	 &	0.088	\\
$\gamma$=0.8	&	2.052	&	0.084	&	48.679	&	1.796	&	0.076	&	22.566	&	2.073	&	0.087	&	4.250	&	2.116	 &	0.092	&	0.463	\\
$\gamma$=0.9	&	\emph{2.239}	&	\textbf{\emph{0.098}}	&	54.367	&	1.906	&	0.076	&	27.719	&	2.113	&	0.088	 &	11.997	&	\textbf{\emph{2.284}}	 &	 \textbf{\emph{0.098}}	&	2.948	\\
$\gamma$=1	&	2.010	&	0.067	&	0	&	1.364	&	0.046	&	0	&	1.784	&	0.067	&	0	&	1.291	&	0.056	&	 0	\\
\hline
\hline
CVaR-CC		&	2.224	&	0.094	&	0.059	&	\textbf{\emph{2.515}}	&	\textbf{\emph{0.108}}	&	0.059	&	2.187	&	 0.095	&	0.044	&	2.130	&	 0.093	&	0.007	
\\
\hline
  \midrule
  & Av ($\cdot 10^{-3}$) & Sh & Time & n. of assets\\
    \cmidrule{1-5}
   CVaR		&	2.168	&	0.094	&	0.033	&	15.3		\\
   Index	&	1.137	&	0.040	&	-	    &	442		\\
     \cmidrule{1-5}
\end{tabular}
\end{sidewaystable}

\noindent From Tables \ref{Table_1}-\ref{Table_4}, except for two cases (EUROSTOXX50 with 5 assets and SP500 with 10 assets) there always exists at least one value of $\gamma$ for which our MILP model outperforms the other competing models, i.e., CVaR-CC, CVaR and the market index. We also note that, as expected, for the SP500 dataset, the GAP values, which are quite high for $p=5,10$, tends to become very small when $p$ increases. In fact, for $p=15,20$ our MILP model produces in-sample solutions that we suppose are very close to the optimal ones, but XPRESS solver was unable to certify their optimality. This would suggest to develop an effective heuristic procedure that combines the good features of location and portfolio selection models in order to obtain optimal or near-optimal in-sample portfolios that will likely produce better out-of-sample performances, in particular, for large size financial dataset.

\vskip 8 pt
\noindent In order to highlight the effectiveness of our approach, we also decided to provide a second set of experiments. In this case, we adopt a period of 104 weeks as in-sample window and 12 weeks (three months) as out-of-sample, with rebalancing allowed every 12 weeks (see also \cite{ManOgrySpe_2007}). For each value of $p$, each model solves overall 204 problems (104 for DJIA, 55 for EUROSTOXX50, 45 for FTSE100). Note that, in this case, we did not consider the SP500 dataset, since we are now interested in comparing our approach with the other programs when our MILP model is able to find optimal (certified) in-sample solutions. In any case, we observed that some preliminary results with the SP500 dataset are in line with the ones reported in Table \ref{Table_4}.
\vskip 8 pt
\noindent Following the results in Tables \ref{Table_5}-\ref{Table_7}, it is clear that, when the MILP model is able to find optimal in-sample solutions, our approach proves to be efficient both from a computational times of view, and always dominates the other methods. To conclude this section, we note that, quite surprisingly, in all the experimental analysis the pure CVaR model is always dominated by the other methods.


\begin{sidewaystable}
\caption{Out-of-sample performances for DJIA with a 12 weeks rebalancing}
\label{Table_5}
\centering
\small
\begin{tabular}{|c| |c|c|c|  |c|c|c|  |c|c|c|  |c|c|c|}
\toprule
   Model& Av-5 ($\cdot 10^{-3}$) & Sh-5  & Time-5 & Av-10 ($\cdot 10^{-3}$) & Sh-10 & Time-10 & Av-15 ($\cdot 10^{-3}$) & Sh-15 & Time-15 & Av-20 ($\cdot 10^{-3}$) & Sh-20 & Time-20 \\
  \hline
  \midrule
$\gamma$=0.1	&	\emph{1.726}	&	\emph{0.077}	&	0.596	&	\emph{1.608}	&	\emph{0.077}	&	0.712	&	1.746	&	 0.084	&	0.952	&	\emph{2.198}	&	\emph{0.104}	 &	0.896	\\
$\gamma$=0.2	&	\emph{1.851}	&	\emph{0.083}	&	0.540	&	\emph{1.616}	&	\emph{0.077}	&	0.681	&	 \emph{1.766}	&	\emph{0.085}	&	0.940	&	\emph{2.145}	&	 \emph{0.100}	&	0.937	\\
$\gamma$=0.3	&	\emph{1.823}	&	\emph{0.081}	&	0.630	&	1.587	&	0.077	&	0.726	&	\emph{1.805}	&	 \emph{0.086}	&	1.244	&	\emph{2.173}	&	\emph{0.102}	 &	1.233	\\
$\gamma$=0.4	&	\emph{1.756}	&	\emph{0.078}	&	0.630	&	1.511	&	0.072	&	0.701	&	\emph{1.786}	&	 \emph{0.085}	&	1.076	&	\emph{2.199}	&	\emph{0.102}	 &	0.875	\\
$\gamma$=0.5	&	\emph{1.691}	&	\emph{0.074}	&	0.686	&	1.529	&	0.072	&	0.798	&	\emph{1.867}	&	 \emph{0.088}	&	0.989	&	\emph{2.200}	&	\emph{0.103}	 &	1.399	\\
$\gamma$=0.6	&	\emph{1.765}	&	\emph{0.077}	&	0.697	&	\emph{1.735}	&	\emph{0.082}	&	0.857	&	 \emph{1.879}	&	\emph{0.090}	&	0.763	&	\emph{2.239}	&	 \emph{0.104}	&	1.233	\\
$\gamma$=0.7	&	1.595	&	0.070	&	0.666	&	\emph{1.664}	&	\emph{0.079}	&	0.768	&	\emph{1.869}	&	 \emph{0.088}	&	0.942	&	\emph{2.193}	&	\emph{0.102}	 &	1.495	\\
$\gamma$=0.8	&	\emph{1.625}	&	\emph{0.071}	&	0.710	&	\emph{1.629}	&	\emph{0.077}	&	0.791	&	 \emph{1.987}	&	\emph{0.094}	&	1.066	&	 \textbf{\emph{2.356}}	&	\textbf{\emph{0.110}}	&	1.189	\\
$\gamma$=0.9	&	\emph{1.855}	&	\emph{0.079}	&	0.683	&	\emph{1.644}	&	\emph{0.076}	&	0.828	&	 \emph{1.877}	&	\emph{0.089}	&	1.089	&	\emph{2.135}	&	 \emph{0.099}	&	1.302	\\
$\gamma$=1	&	\textbf{\emph{2.103}}	&	\textbf{\emph{0.088}}	&	0.101	&	\textbf{\emph{2.127}}	&	\textbf{\emph{0.098}}	 &	0.154	&	\textbf{\emph{2.158}}	&	 \textbf{\emph{0.101}}	&	0.203	&	\emph{2.236}	&	\emph{0.103}	&	0.118	 \\
\hline
\hline
CVaR-CC		&	1.530	&	0.070	&	0.276	&	1.499	&	0.071	&	0.137	&	1.752	&	0.083	&	0.300	&	2.088	&	 0.098	&	0.371	
\\
  \hline
  \midrule
  & Av ($\cdot 10^{-3}$) & Sh & Time & n. of assets\\
    \cmidrule{1-5}
   CVaR		&	1.518	&	0.071	&	0.003	&	8.7		\\
   Index	&	1.608	&	0.068	&	-	    &	28		\\
     \cmidrule{1-5}
\end{tabular}
\end{sidewaystable}

\begin{sidewaystable}
\caption{Out-of-sample performances for EUROSTOXX50 with a 12 weeks rebalancing}
\label{Table_6}
\centering
\small
\begin{tabular}{|c| |c|c|c|  |c|c|c|  |c|c|c|  |c|c|c|}
\toprule
   Model& Av-5 ($\cdot 10^{-3}$) & Sh-5  & Time-5 & Av-10 ($\cdot 10^{-3}$) & Sh-10 & Time-10 & Av-15 ($\cdot 10^{-3}$) & Sh-15 & Time-15 & Av-20 ($\cdot 10^{-3}$) & Sh-20 & Time-20 \\
  \hline
  \midrule
$\gamma$=0.1	&	\emph{1.975}	&	\emph{0.081}	&	1.314	&	\emph{1.816}	&	\emph{0.075}	&	1.418	&	 \emph{1.851}	&	\emph{0.077}	&	2.031	&	1.928	&	0.080	 &	2.022	\\
$\gamma$=0.2	&	\emph{2.004}	&	\emph{0.084}	&	1.490	&	\emph{1.808}	&	\emph{0.075}	&	1.297	&	 \emph{1.844}	&	\emph{0.077}	&	1.529	&	\emph{1.971}	&	 \emph{0.082}	&	2.510	\\
$\gamma$=0.3	&	\emph{2.102}	&	\emph{0.086}	&	1.675	&	\emph{1.821}	&	\emph{0.075}	&	1.712	&	 \emph{1.917}	&	\emph{0.080}	&	2.247	&	1.941	&	0.080	 &	2.066	\\
$\gamma$=0.4	&	\emph{2.117}	&	\emph{0.087}	&	1.652	&	\emph{1.760}	&	\emph{0.073}	&	1.655	&	 \emph{1.899}	&	\emph{0.079}	&	2.105	&	1.929	&	0.080	 &	2.654	\\
$\gamma$=0.5	&	\emph{2.224}	&	\textbf{\emph{0.092}}	&	1.736	&	\emph{1.779}	&	\emph{0.073}	&	1.601	&	 \emph{1.887}	&	\emph{0.078}	&	2.295	&	 \emph{2.045}	&	\emph{0.085}	&	3.068	\\
$\gamma$=0.6	&	\emph{2.061}	&	\emph{0.085}	&	1.930	&	\emph{1.866}	&	\emph{0.077}	&	1.792	&	 \emph{1.896}	&	\emph{0.078}	&	2.758	&	 \textbf{\emph{2.074}}	&	\textbf{\emph{0.086}}	&	2.693	\\
$\gamma$=0.7	&	\emph{1.921}	&	\emph{0.080}	&	2.065	&	1.794	&	0.074	&	2.407	&	\emph{1.951}	&	 \emph{0.080}	&	3.260	&	\emph{2.060}	&	\emph{0.085}	 &	3.665	\\
$\gamma$=0.8	&	\emph{2.069}	&	\emph{0.084}	&	1.930	&	\emph{1.931}	&	\emph{0.079}	&	2.414	&	 \emph{2.021}	&	\emph{0.083}	&	3.295	&	\emph{2.061}	&	 \emph{0.085}	&	4.081	\\
$\gamma$=0.9	&	1.826	&	0.072	&	1.933	&	\emph{1.983}	&	\emph{0.082}	&	2.320	&	\textbf{\emph{2.106}}	&	 \textbf{\emph{0.086}}	&	4.195	&	\emph{2.061}	&	 \emph{0.084}	&	4.184	\\
$\gamma$=1	&	\textbf{\emph{2.228}}	&	\emph{0.077}	&	0.368	&	\textbf{\emph{2.207}}	&	\textbf{\emph{0.090}}	&	 0.272	&	\emph{1.993}	&	\emph{0.079}	&	0.211	&	 \emph{2.069}	&	\emph{0.083}	&	0.216	\\
\hline
\hline
CVaR-CC		&	1.863	&	0.076	&	0.162	&	1.796	&	0.074	&	0.104	&	1.804	&	0.075	&	0.148	&	1.964	&	 0.082	&	0.202	
\\
  \hline
  \midrule
  & Av ($\cdot 10^{-3}$) & Sh & Time & n. of assets\\
    \cmidrule{1-5}
   CVaR		&	1.745	&	0.071	&	0.003	&	6.9		\\
   Index	&	0.712	&	0.024	&	-	    &	49		\\
     \cmidrule{1-5}
\end{tabular}
\end{sidewaystable}

\begin{sidewaystable}
\caption{Out-of-sample performances for FTSE100 with a 12 weeks rebalancing}
\label{Table_7}
\centering
\small
\begin{tabular}{|c| |c|c|c|  |c|c|c|  |c|c|c|  |c|c|c|}
\toprule
   Model& Av-5 ($\cdot 10^{-3}$) & Sh-5  & Time-5 & Av-10 ($\cdot 10^{-3}$) & Sh-10 & Time-10 & Av-15 ($\cdot 10^{-3}$) & Sh-15 & Time-15 & Av-20 ($\cdot 10^{-3}$) & Sh-20 & Time-20 \\
  \hline
  \midrule
$\gamma$=0.1	&	\emph{2.771}	&	\emph{0.123}	&	13.531	&	2.180	&	0.103	&	4.895	&	2.275	&	0.109	&	 2.637	&	2.207	&	0.108	&	3.033	\\
$\gamma$=0.2	&	\emph{2.670}	&	\emph{0.119}	&	19.999	&	\emph{2.322}	&	\textbf{\emph{0.110}}	&	5.695	&	 \emph{2.325}	&	\textbf{\emph{0.111}}	&	2.486	&	 2.257	&	0.109	&	3.287	\\
$\gamma$=0.3	&	2.290	&	0.099	&	20.967	&	2.138	&	0.101	&	7.448	&	2.236	&	0.107	&	2.898	&	2.198	 &	0.107	&	3.274	\\
$\gamma$=0.4	&	2.197	&	0.097	&	24.230	&	2.125	&	0.100	&	8.006	&	2.195	&	0.104	&	2.864	&	2.183	 &	0.105	&	2.909	\\
$\gamma$=0.5	&	2.495	&	0.112	&	24.582	&	2.047	&	0.096	&	7.776	&	2.177	&	0.103	&	3.171	&	2.120	 &	0.102	&	3.062	\\
$\gamma$=0.6	&	2.013	&	0.091	&	26.766	&	2.131	&	0.100	&	8.964	&	2.175	&	0.103	&	3.318	&	2.151	 &	0.104	&	3.257	\\
$\gamma$=0.7	&	\emph{2.626}	&	\emph{0.117}	&	24.267	&	2.190	&	0.103	&	9.490	&	2.068	&	0.098	&	 3.841	&	2.170	&	0.104	&	2.976	\\
$\gamma$=0.8	&	\textbf{\emph{3.025}}	&	\textbf{\emph{0.131}}	&	17.133	&	2.283	&	0.107	&	9.134	&	2.128	&	 0.101	&	4.018	&	2.187	&	0.104	&	3.770	\\
$\gamma$=0.9	&	1.459	&	0.059	&	10.543	&	\textbf{2.341}	&	0.107	&	8.135	&	2.109	&	0.099	&	4.700	&	 2.221	&	0.104	&	3.893	\\
$\gamma$=1	&	0.876	&	0.029	&	1.803	&	1.529	&	0.065	&	2.025	&	\textbf{2.452}	&	0.107	&	1.311	&	 \textbf{\emph{2.560}}	&	\textbf{\emph{0.118}}	&	1.099	 \\
\hline
\hline
CVaR-CC		&	2.417	&	0.113	&	0.446	&	2.235	&	0.106	&	0.337	&	2.273	&	0.109	&	0.273	&	2.270	&	 0.111	&	0.316	
\\
  \hline
  \midrule
  & Av ($\cdot 10^{-3}$) & Sh & Time & n. of assets\\
    \cmidrule{1-5}
   CVaR		&	2.315	&	0.109	&	0.009	&	11		\\
   Index	&	0.567	&	0.021	&	-	    &	83		\\
     \cmidrule{1-5}
\end{tabular}
\end{sidewaystable}

\newpage

\section{Conclusion}
In this paper we propose a novel framework for portfolio selection that combines the specific features of a clustering and a portfolio optimization techniques through the global solution of a hard Mixed-Integer Linear Programming problem. The idea of our approach is to overcome the classical two phases approach characterized by two distinct steps: cluster first and then selection. We show that our method is quite general since other risk measures as well as correlation measures can be adopted in our framework. The resulting MILP program is, in fact, \emph{portable} in the sense that in model (\ref{D1})-(\ref{D7}), only equations (\ref{D1})-(\ref{D4}) depend on the specific risk measure adopted. Actually, constraints (\ref{D5})-(\ref{D7}) are independent of such risk measure and they can be used in the formulation of MILP programs based on other measures (e.g., Mean Absolute Deviation, Gini's index, minimax objective function etc...). Our model was tested on real financial dataset, compared to some benchmark models, and found to give good results in terms of realized profit. We also point out that the MILP program (\ref{D1})-(\ref{D7}) can be efficiently solved at least for moderate sized problems.

To conclude, the results reported in this paper are encouraging and there is room for improving the optimization phase by providing an ad hoc heuristic approach for solving large size dataset. This is, in fact, one of our future lines of research.


\newpage

\begin{thebibliography}{99}

\bibitem{ArtDelEberHea_1999}
Artzner, P., Delbaen, F., Eber, J. M., Heath, D.: Coherent Measures of Risk. Mathematical Finance 9, 203--228 (1999)

\bibitem{BellDiber_2017}
Bellini, F., Di Bernardino, E.: Risk management with expectiles. The European Journal of Finance 23, 487--506 (2017)

\bibitem{Benati_2008}
Benati, S.: Categorical data fuzzy clustering: an analysis of local search heuristics. Computers $\&$ Operations Research 35, 766--775 (2008)

\bibitem{BenGar_2014}
Benati, S., Garc\'{i}a, S.: A mixed integer linear model for clustering with variable selection. Computers $\&$ Operations Research 43, 280--285 (2014)

\bibitem{BeGaPu_2018}
Benati, S., Garc\'{i}a, S., Puerto, J.: Mixed integer linear programming and heuristic methods for feature selection in clustering. Journal of the Operational Research Society 69, 1379--1395 (2018)

\bibitem{BerBru_2014}
Beraldi, P., Bruni, M.E.: A clustering approach for scenario tree reduction: an application to a stochastic programming portfolio optimization problem. TOP 22, 934--949 (2014)

\bibitem{BogButeShiTruLaf_2014}
Boginski, V., Butenko, S., Shirokikh, O., Trukhanov, S., Lafuente, J.G.: A network-based data mining approach to portfolio selection via weighted clique relaxations. Annals of Operations Research 216, 23--34 (2014)

\bibitem{BruCeScoTar_2017}
Bruni, R., Cesarone, F., Scozzari, A., Tardella, A.: On exact and approximate stochastic dominance strategies for portfolio selection. European Journal of Operational Research 259, 322--329 (2017)

\bibitem{BruCeScoTar_2016}
Bruni, R., Cesarone, F., Scozzari, A., Tardella, A.: Real-world datasets for portfolio selection and solutions of some stochastic dominance portfolio models. Data in Brief 8, 858--862 (2016)

\bibitem{CeScoTar_2013}
Cesarone, F., Scozzari, A., Tardella, A.: A new method for mean-variance portfolio optimization with cardinality constraints. Annals of Operations Research 205, 213--234 (2013)

\bibitem{CeTar2017}
Cesarone, F., Tardella, A.: Equal Risk Bounding is better than Risk Parity for portfolio selection. Journal of Global Optimization 68, 439--461 (2017)

\bibitem{CeScoTar_2015}
Cesarone, F., Scozzari, A., Tardella, A.: Linear vs. quadratic portfolio selection models with hard real world constraints. Computational Management Science 12, 345--370 (2015)

\bibitem{ChangMeadeBeasley_2000}
Chang, T.J., Meade, N., Beasley, J.E., Sharaiha, Y.M.: Heuristics for cardinality constrained portfolio optimisation. Computers $\&$ Operations Research 27, 1271--1302 (2000)

\bibitem{ChopraZiemba_2013}
Chopra, V. K., Ziemba, W. T.: The effect of errors in means, variances, and covariances on optimal portfolio choice. In Handbook of the Fundamentals of Financial Decision Making: Part I, pp. 365-373 (2013)

\bibitem{ClemGraHitaj2019}
Clemente, G. P., Grassi, R., Hitaj, A.: Asset allocation: new evidence through network approaches. Annals of Operations Research. https://doi.org/10.1007/s10479-019-03136-y (2019)

\bibitem{CornPerl_1984}
Corneil, D.G., Perl, Y.: Clustering and domination in perfect graphs. Discrete Applied Mathematics 9, 27--39 (1984)

\bibitem{CornStewart_1990}
Corneil, D.G., Stewart, L.K.: Dominating sets in perfect graphs. Discrete Mathematics 86, 145--164 (1990)

\bibitem{DeMiGar2007}
DeMiguel, V., Garlappi, L., Uppal, R.: Optimal versus naive diversification: How inefficient is the $1/n$ portfolio strategy? The Review of Financial studies 22, 1915--1953 (2007)

\bibitem{EltGruSpi2006}
Elton, E.J., Gruber, M.J., Spitzer, J.: Improved estimates of correlation coefficients and their impact on optimum portfolios. European Financial Management 12, 303--318 (2006)

\bibitem{Ehrgott_2005}
Ehrgott, M.: Multicriteria Optimization. Vol. 491. Springer Science $\&$ Business Media, Berlin Heidelberg (2005)

\bibitem{Ehrgott_2006}
Ehrgott, M.: A discussion of scalarization techniques for multiple objective integer programming. Annals of Operations Research 147, 343--360 (2006)

\bibitem{GarLabMar_2011}
Garc\'{i}a, S., Labb\'{e}, M., Mar\'{i}n, A.: Solving large $p$-median problems with a radius formulation. INFORMS Journal on Computing 23, 546--556 (2011)

\bibitem{GareyJohnson_1979}
Garey, M.R., Johnson, D.S.: Computers and intractability: A guide to the theory of NP-Completeness. W.H. Freeman and Company, New York (1979)

\bibitem{Hansen_et_al_2009}
Hansen, P., Brimberg, J., Urosevic, D., Mladenovic, N.: Solving large p-median clustering problems by primal-dual variable neighborhood search. Data Mining and Knowledge Discovery 19, 351--375 (2009)

\bibitem{JegTit_2001}
Jegadeesh, N., Titman, S.: Profitability of momentum strategies: An evaluation of alternative explanations. The Journal of Finance 56, 699--20 (2001)

\bibitem{Job_Kork1980}
Jobson, J. D., Korkie, B.: Estimation for Markowitz efficient portfolios. Journal of the American Statistical Association 75, 544-554 (1980)

\bibitem{KarHak_1979}
Kariv, O., Hakimi, L.: An Algorithmic Approach to Network Location Problems. II: The p-Medians. SIAM Journal on Applied Mathematics 37, 539--560 (1979)

\bibitem{KonnoYamazaki_1991}
Konno, H., Yamazaki, H.: Mean-absolute deviation portfolio optimization model and its applications to Tokyo stock market. Management science, 37, 519--531 (1991).

\bibitem{KulKerKas_2002}
Kullmann, L., Kertesz, J., Kaski, K.: Time-dependent cross-correlations between different stock returns: A directed network of influence. Physical Review E 66, 026125 (2002)

\bibitem{KulKerMant_2000}
Kullmann, L., Kertesz J., Mantegna, R.: Identification of clusters of companies in stock indices via Potts super-paramagnetic transitions. Physica A: Statistical Mechanics and its Applications 287, 412--419 (2000)

\bibitem{MaiRonTel_2010}
Maillard, S., Roncalli, T., Telietche, J.: The properties of equally weighted risk contribution portfolios. The Journal of Portfolio Management 36, 60--70.

\bibitem{ManOgrySpe_2007}
Mansini, R., Ogryczak, W., Speranza, M. G.: Conditional value at risk and related linear programming models for portfolio optimization. Annals of operations research 152, 227--256  (2007)

\bibitem{ManOgrySpe_2015}
Mansini, R., Ogryczak, W., Speranza, M.G.: Linear and Mixed Integer Programming for Portfolio Optimization, Springer, Switzerland (2015)

\bibitem{Mantegna_1999}
Mantegna, R.N.: Hierarchical structure in financial markets. The European Physical Journal B-Condensed Matter and Complex Systems
11, 193--197 (1999)

\bibitem{Markowitz1952}
Markowitz, H.: Portfolio selection. The Journal of Finance 7, 77--91 (1952)

\bibitem{Merton_1980}
Merton, R.C.: On estimating the expected return on the market: An exploratory investigation. Journal of Financial Economics 8, 323--361 (1980)

\bibitem{MirchFran_1990}
Mirchandani, P.B., Francis, R.L.: Discrete location theory, Wiley-Interscience Series in Discrte Mathematics and Optimization, New York (1990)

\bibitem{NgHan_1994}
Ng, R.T., Han, J.: Efficient and effective clustering methods for spatial data mining, in Proceedings of the 20th International Conference on Very Large Data Bases, VLDB '94, San Francisco, CA, USA, 1994, Morgan Kaufmann Publishers Inc., pp. 144-155.

\bibitem{Onnela_et_al_2004}
J.-P. Onnela, J. P., Kaski, K., Kertesz J.: Clustering and information in correlation based financial networks. The European Physical Journal B 38, 353--362 (2004)

\bibitem{RockafellarUryasev_2002}
Rockafellar, R. T., Uryasev, S.: Conditional value-at-risk for general loss distributions. Journal of banking $\&$ finance 26, 1443--1471 (2002)

\bibitem{Sharpe1966}
Sharpe, W.F.: Mutual fund performance. Journal of Business 39, 119--138 (1966)

\bibitem{Sharpe1994}
Sharpe, W.F.: The sharpe ratio. The Journal of Portfolio Management 21, 49--58 (1996)

\bibitem{Simaan_1997}
Simaan, Y.: Estimation risk in portfolio selection: The Mean Variance Model and the Mean-Absolute Deviation model. Management Science 43, 1437--1446 (1997)


\bibitem{Tola_et_al_2008}
Tola, V., Lillo, F., Gallegati, M., Mantegna, R.N.: Cluster analysis for portfolio optimization. Journal of Economic Dynamics and Control 32, 235--258 (2008)

\bibitem{WattsStrogatz_1998}
Watts, D.J., Strogatz, S.H.: Collective dynamics of small-worldnetworks. Nature 393(6684), 440 (1998)




\end{thebibliography}
\end{document}